\documentclass{article}

\usepackage{amssymb}
\usepackage{amsmath}
\usepackage{amsfonts}
\usepackage{amsthm}
\usepackage{algpseudocode}
\usepackage{algorithm}
\usepackage{graphicx}
\usepackage{geometry}
\usepackage{color}
\usepackage{natbib}
\usepackage{setspace}
\usepackage{overpic}
\usepackage{blkarray}
\usepackage{authblk}

\normalfont\upshape

\newcommand{\yobs}{y^\mathsf{obs}}
\newcommand{\intd}{\mathsf{d}}

\newcommand{\ABC}[2]{p\left( #2 \mid y \in #1 \right) }

\newcommand{\indicator}[1]{\mathbb{I}\left( #1 \right)}
\newcommand{\dist}[2]{\rho ( #1, #2 ) }
\newcommand{\normal}[1]{\mathsf{Normal} \left( #1 \right) }
\newcommand{\unif}[2]{\mathsf{Unif}_{#1} \left( #2 \right) }

\newcommand{\prob}[1]{\mathsf{P} \left( #1 \right) }
\newcommand{\acceptancereg}[2]{\Omega_{\left[#1, #2\right)}}

\newcommand{\KL}[2]{\text{D}_{\text{KL}}\left( #1 || #2 \right)}

\newcommand{\jointproposal}[1]{q_{#1}}
\newcommand{\optimaljointproposal}[1]{\jointproposal{#1}^\prime}

\newtheorem{exmp}{Example}

\title{Stratified distance space improves the efficiency of sequential samplers for approximate Bayesian computation}

\date{}

\author[1,*]{Henri Pesonen}
\author[2,3,4]{Jukka Corander}
\affil[1]{Oslo Center for Biostatistics and Epidemiology, Oslo University Hospital, Oslo, Norway}
\affil[2]{Department of Biostatistics, University of Oslo, Oslo, Norway}
\affil[3]{Parasites and Microbes, Wellcome Sanger Institute, Hinxton, Cambridgeshire, UK}
\affil[4]{Helsinki Institute for Information Technology HIIT, Department of Mathematics and Statistics, University of Helsinki Helsinki, Finland}
\affil[*]{Corresponding author: henri.e.pesonen@medisin.uio.no}

\begin{document}

\maketitle

\begin{abstract}
    Approximate Bayesian computation (ABC) methods are standard tools for inferring parameters of complex models when the likelihood function is analytically intractable. A popular approach to improving the poor acceptance rate of the basic rejection sampling ABC algorithm is to use sequential Monte Carlo (ABC SMC) to produce a sequence of proposal distributions adapting towards the posterior, instead of generating values from the prior distribution of the model parameters. Proposal distribution for the subsequent iteration is typically obtained from a weighted set of samples, often called particles, of the current iteration of this sequence. Current methods for constructing these proposal distributions treat all the particles equivalently, regardless of the corresponding value generated by the sampler, which may lead to inefficiency when propagating the information across iterations of the algorithm. To improve sampler efficiency, a modified approach called stratified distance ABC SMC is introduced. The algorithm stratifies particles based on their distance between the corresponding synthetic and observed data, and then constructs distinct proposal distributions for all the strata. Taking into account the distribution of distances across the particle space leads to substantially improved acceptance rate of the rejection sampling. It is shown that further efficiency could be gained by using a newly proposed stopping rule for the sequential process based on the stratified posterior samples and these advances are demonstrated by several examples.
\end{abstract}

\section{Introduction}

Likelihood-free inference (LFI) methods have become standard tools for statistical inference when the likelihood functions are not available in closed form but we are able to simulate from the model given permissible parameter values of our choice \citep{cranmer2019}. Simulation-based inference is geared to be one of the cornerstones of many intelligence-based systems \citep{lavin2021}. Approximate Bayesian computation (ABC) methods were within the first LFI methods and even the most basic rejection sampling ABC method can arguably still be relevant due to its simplicity and easy applicability when forward simulation from a model is possible \citep{PitchardEtAl1999, beaumont2002approximate, tanaka2006}. ABC methods continue to be actively developed and its advanced extensions are being successfully used in new applications as LFI is gaining popularity in different research fields \citep{sisson2018, lintusaari2017, martin2021, pesonen2022, engebretsen2023}. 

Within the first extensions of ABC methods were sequential Monte Carlo ABC methods \citep{sisson:fan:tanaka:2007, toni2009, DelMoralEtAl2012}. They improve on the efficiency of rejection sampling based ABC methods by introducing a sequence of importance sampling distributions to improve acceptance rate of the proposed simulated samples. ABC SMC methods are actively developed and many proposals have been made to improve the accuracy and efficiency by introducing improved importance sampling distribution generation strategies or adaptive threshold selection \citep{Silk2013, filippi2013, bonassi2015, simola2020adaptive}. Other strategies to improve sequential sampling is to incorporate either multilevel Monte Carlo strategies \citep{jasra2019} or delayed acceptance sampling  \citep{everitt2021} where computationally faster lower level approximate models can be used to reduce the number of generated data points from the expensive accurate models.
Diverging from the sampling based ABC methods, synthetic likelihood (SL) methods \citep{wood2010, price2018} directly find an estimate for the unknown likelihood function using a set of simulated samples. The development of both approaches, ABC and SL have been accelerated by the introduction of powerful machine learning techniques. \citet{gutmann2016bayesian} introduced an efficient Bayesian optimisation-based approach Bayesian optimisation for likelihood-free inference (BOLFI) to find a surrogate for the discrepancy metric as a function of the unknown parameters. \citet{lueckmann2018} and \citet{papamakarios2019} introduced neural network-based flexible models that can be used as a surrogates for the likelihood function or directly for the posterior distribution.

In this study we introduce a new variant to the ABC SMC methods where we use the partitioning of the ABC acceptance regions to find better proposal distributions for the importance sampling and improve the posterior approximation. 
In Section \ref{sec:overview} we discuss ABC SMC methods and present the theory for optimal selection of particular type of propagation distributions. We present the new inference methodology in Section \ref{sec:stratified_kernels} with a novel early stopping rule in Section \ref{sec:stopping_rule}. In Section \ref{sec:examples} we present the results of the numerical experiments and conclude the article with discussion in Section \ref{sec:conclusion}.


\section{Overview of ABC SMC methods}\label{sec:overview}

ABC methods are based on approximations to the underlying models that enable us to draw a sample from an approximation to posterior distribution $p(\theta \mid \yobs)$. 
Instead of using the likelihood we draw a joint sample of the parameters $\theta_i$ (called particles) and observations $y_i$ using the prior $p(\theta)$ and the simulator model $f(y \mid \theta)$ and retain $N$ tuples $(\theta_i ,y_i, w_i)$ where the simulated observation is sufficiently similar to the actual observation $\yobs$.
Similarity is determined via a discrepancy measure $\dist{y}{\yobs}$ and an error tolerance threshold $\epsilon$. Discrepancy measure and the threshold define an \textit{ABC acceptance region} of simulated observations in the observation space.
We define $\epsilon_0 := 0$, ABC acceptance region as $\acceptancereg{\epsilon_a}{\epsilon_b} := \{y: \epsilon_a \leq \dist{y}{\yobs} < \epsilon_b \}$ and the ABC posterior as
\begin{align}
    p(\theta \mid \yobs) \approx \ABC{\acceptancereg{0}{\epsilon}}{ \theta} &= \frac{p(y \in \acceptancereg{0}{\epsilon} | \theta)p(\theta)}{\int p(y \in \acceptancereg{0}{\epsilon} | \theta)p(\theta) \intd \theta}.
\end{align}
The proxy for the likelihood is
\begin{align}
    p(y \in \acceptancereg{0}{\epsilon} \mid \theta) =  \int f(y \mid \theta) \indicator{y \in \acceptancereg{0}{\epsilon}} \intd y,
\end{align}
where $\indicator{\cdot}$ is the indicator function. Often we use summary statistics $S(y)$ and $S(\yobs)$ instead of the original data, but here we assume that data can be either the non-transformed or summarised and will not consider the information loss related to the transformation. 

This basic ABC method is referred to as the \textit{rejection ABC}. Arguably the main problem with it is the potentially poor acceptance rate associated with reasonable thresholds $\epsilon$ which is generally caused by the use of prior as a proposal distribution. 
To make ABC algorithm more efficient, importance sampling based sequential Monte Carlo techniques \citep{chopin2020} have been applied in ABC. The resulting sequential Monte Carlo ABC (ABC SMC) techniques were introduced to produce a sequentially improving set of proposal distributions. To improve the ABC posterior, the algorithm uses a decreasing sequence of acceptance thresholds $\epsilon_{1}, \ldots, \epsilon_T$ and random walk proposal kernels $K_t(\theta^{(t+1)} \mid \theta^{(t)})$. Kernels are used to propagate a set of weighted samples $\left\{\left(\theta_i^{(t)}, w_i^{(t)}, y_i^{(t)} \right)\right\}_{i=1}^N$
that represent the current posterior distribution at iteration $t$ to $t+1$. The ABC posterior sample of the current iteration $t$ is often used to set the parameters of the kernel. Although other choices are possible, most often the kernels are zero-centered Gaussian distributions where the covariance matrices $\Sigma_t$ are empirically calculated from a set of weighted ABC posterior samples
\begin{align}\label{eq:gaussian_kernel}
K_t(\theta^{(t+1)} \mid \theta^{(t)}) &=  \normal{\theta^{(t+1)} \mid \theta^{(t)},  \Sigma_t}.
\end{align}

The ABC SMC joint proposal distribution for the tranformation $t \rightarrow t+1$  is of the form
\begin{align}\label{eq:ABC_proposal}
    & \jointproposal{t}\left(\theta^{{(t)}}, \theta^{(t+1)}\right) \\\nonumber & =\frac{\ABC{\acceptancereg{0}{\epsilon_t}}{\theta^{(t)}}K_t(\theta^{(t+1)} | \theta^{(t)})  \prob{y^\prime \in \acceptancereg{0}{\epsilon_{t+1}} \mid \theta^{(t+1)}}}{\int \int \ABC{\acceptancereg{0}{\epsilon_t}}{\theta^{(t)}}K_t(\theta^{(t+1)} | \theta^{(t)}) \prob{y^\prime \in \acceptancereg{0}{\epsilon_{t+1}} \mid \theta^{(t+1)}} \intd \theta^{(t+1)}\intd \theta^{(t)}}.
\end{align}

\citet{filippi2013} derived an optimal Gaussian kernel \eqref{eq:gaussian_kernel} by minimizing the Kullback-Leibler (KL) discrepancy between \eqref{eq:ABC_proposal} and
\begin{align}\label{eq:optimal_product}
\optimaljointproposal{t}\left(\theta^{{(t)}}, \theta^{(t+1)}\right) = 
p\left(\theta^{(t)} \mid y \in \Omega_{[0, \epsilon_t)}\right)p\left(\theta^{(t+1)} \mid y^\prime \in \Omega_{[0, \epsilon_{t+1}) }\right)
\end{align}
that is defined as
\begin{align}
    \KL{\jointproposal{t}}{\optimaljointproposal{t}} &= \int \int \log \left( \frac{\optimaljointproposal{t}\left(\theta^{{(t)}}, \theta^{(t+1)}\right)}{\jointproposal{t}\left(\theta^{{(t)}}, \theta^{(t+1)}\right)} \right) \optimaljointproposal{t}\left(\theta^{{(t)}}, \theta^{(t+1)}\right) \intd \theta^{(t)} \intd \theta^{(t+1)}.
\end{align}

Given the assumptions \eqref{eq:gaussian_kernel} and \eqref{eq:optimal_product} we can derive an optimal covariance for the  Gaussian kernel
\begin{align}\label{eq:optimal_kernel_covariance}
    \Sigma^{(t)} = \int \! \! \int & (\theta^{(t)} \!-\! \theta^{(t+1)})(\theta^{(t)} \!-\! \theta^{(t+1)})^T \times \nonumber \\ & \times \ABC{\acceptancereg{0}{\epsilon_{t}}}{\theta^{(t)}} p\left(\theta^{(t+1)} \mid y^\prime \in \Omega_{ [0, \epsilon_{t+1}) }\right) \intd \theta^{(t)} \intd \theta^{(t+1)},
\end{align}
that can be approximated using the ABC posterior sample 
by applying the threshold $\epsilon_{t+1}$ to select a re-weighted $N^\prime$ sized subset of it to represent $\ABC{\acceptancereg{0}{\epsilon_{t+1}}}{\theta^{(t+1)}}$
\begin{align}\label{eq:abc_sample}
    \left\{ \left( \theta^{(t+1)}_{I(i)}, \widetilde{w}^{(t+1)}_{I(i)}, y^{(t+1)}_{I(i)} \right)  \right\}_{i=1}^{N^\prime} = \left\{ \left(\theta_{I(i)}^{(t)}, \frac{w_{I(i)}^{(t)}}{\sum_{j=1}^{N^\prime}w_{I(j)}^{(t)}}, y_{I(i)}^{(t)}\right) : y_{I(i)} \in \acceptancereg{0}{\epsilon_{t+1}}\right\}_{i=1}^{N^\prime},
\end{align}
where the subset of indices $\{I(i)\}_{i=1}^{N^\prime} \subset \{1, 2, \ldots,  N \}$ indicate samples for which $y_{I(i)} \in \acceptancereg{0}{\epsilon_{t+1}}$.

After obtaining the random walk proposal, we use it to generate candidates $\theta^\prime$ by propagating randomly selected particles $\theta^{(t)}_i$ with sampling probabilities $w^{(t)}_i$  from ABC posterior using the kernel $\theta^\prime \thicksim K_t\left( \theta^{(t+1)} | \theta^{(t)}_i\right)$ and generating new synthetic observations $y^\prime \thicksim p(y \mid \theta^\prime)$. The new synthetic data are then compared against the observed data to determine whether or not it is within the acceptance regions of the next iteration $y^\prime \in \acceptancereg{0}{\epsilon_{t+1}}$ and the sampling is repeated until a sample set of size $N$ is obtained. The re-weighting follows the general importance sampling procedure
\begin{align}\label{eq:abcsmcweight}
    w^{(t+1)}_i \propto \frac{p\left(\theta^{(t+1)}_i\right)}{\sum_{j=1}^N w_j^{(t)} K_{t}\left(\theta^{(t+1)}_i |\theta^{(t)}_j \right)},
\end{align}
where  $p(\cdot)$ is the prior of the parameters and the importance sampling distribution $\sum_{j=1}^N w_j^{(t)} K_{t}\left(\theta^{(t+1)} |\theta^{(t)}_j \right)$ is empirically estimated from the weighted sample as a Gaussian mixture. 

\citet{filippi2013} introduced a locally optimal Gaussian kernel with analogous covariance to \eqref{eq:optimal_kernel_covariance} by optimising for the covariance for each particle $\theta^{(t)}_i$ separately, i.e.~setting effectively $\ABC{\acceptancereg{0}{\epsilon_t}}{\theta} = \delta(\theta - \theta_i^{(t)})$, where $\delta(\cdot)$ is the delta function.
This leads to a kernel covariance
\begin{align}\label{eq:locally_optimal_cov}
    \Sigma_{i}^{(t)} &= \int (\theta^{(t)}_i - \theta^{(t+1)})(\theta^{(t)}_i - \theta^{(t+1)})^T  \ABC{\acceptancereg{0}{\epsilon_{t+1}}}{\theta^{(t+1)}} \intd \theta^{(t+1)},
\end{align}
which can again be approximated based on the sample \eqref{eq:abc_sample}.

\section{Stratified sampling ABC SMC}\label{sec:stratified_kernels}

In most cases ABC SMC is superior to the basic rejection ABC algorithm, especially when the prior distribution of the parameters is far off from the posterior distribution. There are multiple approaches for improving the standard ABC SMC, by choosing different proposal and target distributions, determining the threshold sequence adaptively or setting rules for stopping the sequential process. Especially in the early iterations of the algorithm, the kernels calculated empirically from the data can have large scale parameters which in turn can result in slower convergence to the approximate posterior distribution that is within our tolerance of error.  
We can investigate ABC SMC more closely by looking at how the particles are accepted and how the weights for the particles are being formed. At $t$th iteration, ABC posterior sample approximation is $\left\{ \left(\theta_i^{(t)}, w_i^{(t)}, y_i^{(t)}\right) \right\}_{i=1}^N$. By using ABC SMC algorithm that was described in Section \ref{sec:overview}, the posterior approximation at $(t+1)$th iteration will be $\left\{ \left( \theta_i^{(t+1)}, w_i^{(t+1)}, y_i^{(t+1)} \right) \right\}_{i=1}^{N_\text{tot}^{(t+1)}}$, where $\theta^{(t+1)}_i\mid \theta_{j}^{(t)} \stackrel{\text{i.i.d}}{\thicksim} \sum_{j=1}^N w_{j}^{(t)} K_t(\theta^{(t+1)} \mid \theta_j^{(t)})$, $N_\text{tot}^{(t+1)} \geq N$ is the total number of simulated samples at $(t+1)$th iteration, and 
\begin{align}
w_i^{(t+1)} \propto  \left\{
\begin{array}{cc} 
\frac{p\left( \theta^{(t+1)}_i \right)}{\sum_{j=1}^N w_j^{(t)}K_t\left( \theta^{(t+1)}_i \mid \theta^{(t)}_j \right)}, & \text{ if } y^{(t+1)}_i \in \acceptancereg{0}{\epsilon_{t+1}},  \left[ \text{ w.p. } \prob{y \in \acceptancereg{0}{\epsilon_{t+1}} \mid \theta^{(t+1)}_i} \right] \\
0, & \text{ if } y^{(t+1)}_i \notin \acceptancereg{0}{\epsilon_{t+1}},  \left[ \text{ w.p. } \prob{y \notin \acceptancereg{0}{\epsilon_{t+1}} \mid \theta^{(t+1)}_i} \right].
\end{array} 
\right. 
\end{align}
Samples are generated until $N$ samples have non-zero weights, so it would be computationally desirable to generate as many samples as possible so that $y^{(t+1)}_i \in \acceptancereg{0}{\epsilon_{t+1}}$, and achieve a large acceptance rate $N/N^{(t+1)}_\text{tot}$. If $\prob{y \in \acceptancereg{0}{\epsilon_{t+1}} \mid \theta}$ would be available, it would be possible to query more samples from the parts of the parameter space that generate non-zero weighted samples with higher probability. Unfortunately this is in practice impossible in any realistic application of ABC. 
The average acceptance probability at $(t+1)$th iteration can be expressed as 
\begin{align}\label{eq:average_acceptance_rate}
    \int \int  \prob{y^\prime \in \acceptancereg{0}{\epsilon_{t+1}} \mid \theta^{(t+1)}} K_t(\theta^{(t+1)} | \theta^{(t)}) \ABC{\acceptancereg{0}{\epsilon_t}}{\theta^{(t)}} \intd \theta^{(t)} \intd \theta^{(t+1)},
\end{align}
which we aim to improve especially at the early iteration rounds $t$ by introducing a novel importance sampling approach that modifies the locally optimal ABC SMC defined by the kernel covariance \eqref{eq:locally_optimal_cov} using a partition of the ABC posterior approximation into components defined by a set of non-overlapping acceptance regions. 
The observation space bounded to the acceptance region at threshold $\epsilon_t$ of ABC SMC can be partitioned based on the threshold sequence
\begin{align}
\acceptancereg{0}{\epsilon_t} = \bigcup_{k=t}^T \acceptancereg{\epsilon_{k+1}}{\epsilon_{k}}, 
\end{align}
where $\epsilon_{T+1} := 0$.
We can use the partitioning and the law of total probability to write the ABC posterior correspondingly as
\begin{align}\label{eq:partitioning}
     \ABC{\acceptancereg{0}{\epsilon_t}}{\theta^{(t)}} & = \sum_{k=t}^T 
     p(\theta^{(t)}\mid y^\prime \in  \acceptancereg{\epsilon_{k+1}}{\epsilon_{k}})  \prob{y^\prime \in \acceptancereg{\epsilon_{k+1}}{\epsilon_{k}} \mid y \in \acceptancereg{0}{\epsilon_t}}.
\end{align}

\begin{exmp}\label{ex:1}
We define a generative model for a single observation from a Gaussian model with mean $\theta$ and variance $1$ with uniform prior for $\theta$
\begin{align}
p(y \mid \theta) = N(y \mid \theta,1), \quad p(\theta) &= \unif{[-6,6]}{\theta}
\end{align}
and observe $\yobs = 0$. Distance measure is the absolute difference between synthetic and observed data. Figure \ref{Fig:posterior_partitioning} illustrates the initial situation where an acceptance threshold sequence $\epsilon_{1:5}$ is selected as $[\infty, 4, 3, 2, 1]$. The first iteration ABC posterior corresponds to the prior and the threshold $\epsilon_1 = \infty$. 
\begin{figure}[ht!]
   \centering
   \includegraphics[trim={0 0 0 0}, clip, width = 0.7\textwidth]{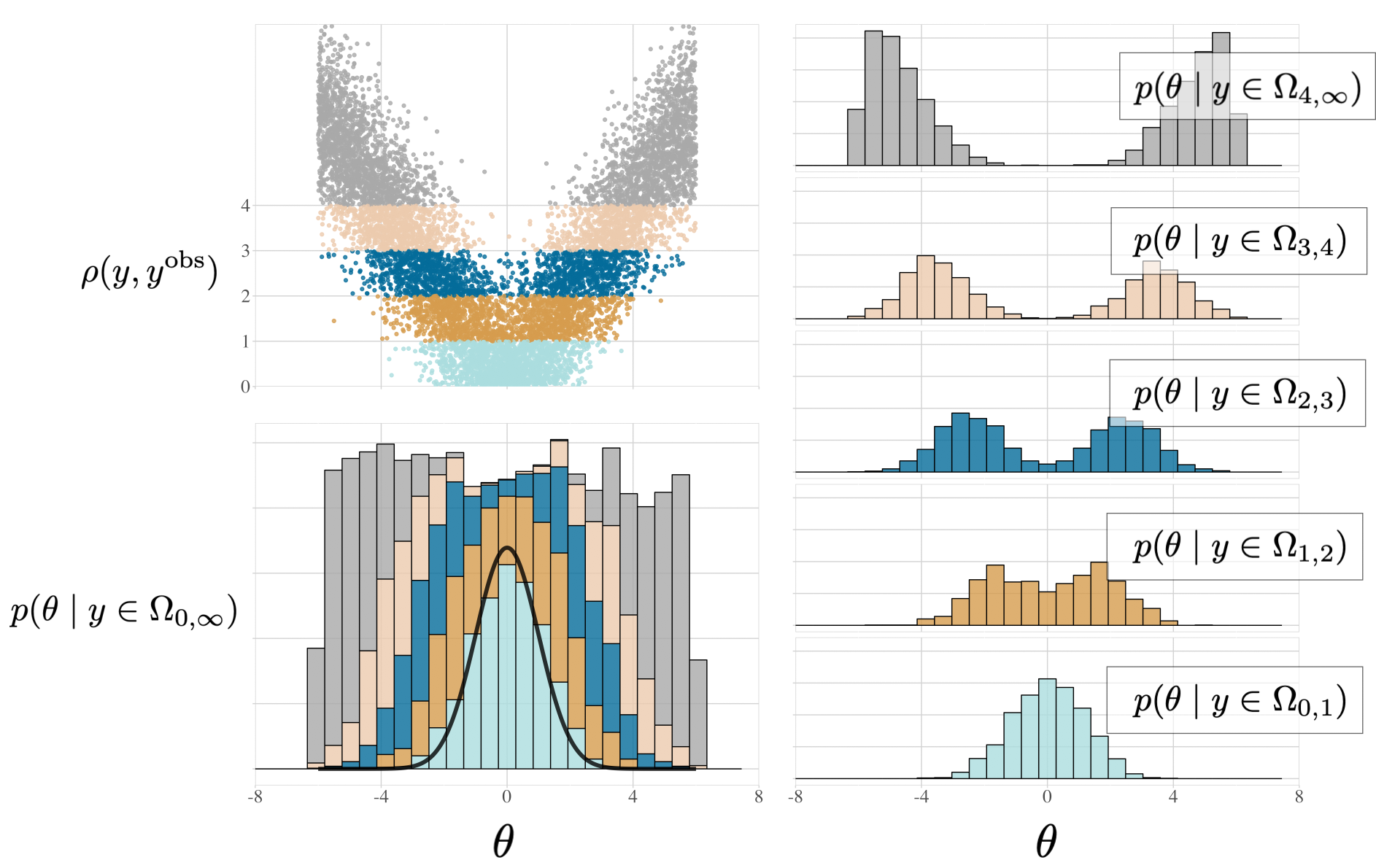}
   \caption{Example 1. Partitioning of the ABC posterior distribution based on the acceptance regions. Here the posterior is actually the prior as the acceptance region $\Omega_{0, \infty}$ is the whole observation space. In the first round of  ABC SMC we'd wish to transfer particles $p(\theta \mid y \in \Omega_{0, \infty})$ into $p(\theta \mid y \in \Omega_{0, 4})$ that is a combination of the four other posterior components.}
   \label{Fig:posterior_partitioning}
\end{figure} 
\end{exmp}

The partitioned posterior motivates the proposed ABC SMC approach. We propose a modification to the locally optimal Gaussian transition kernel by removing the dependence on iteration $t$, and instead define it for each posterior component conditioned on the events of observations belonging to the acceptance regions $\acceptancereg{\epsilon_{k+1}}{\epsilon_{k}}$.  Given the set of predefined acceptance regions and the ABC posterior sample at $i$th iteration, we first define a locally optimal transition kernel similar to \eqref{eq:locally_optimal_cov} by taking into account which partitioned acceptance region band a particle resides in and then balance the particle selection probability by using approximations to the weights in \eqref{eq:partitioning} that we get as a by-product of previous iterations of ABC SMC. 

Let $(\theta_i^{(t)}, y_i^{(t)}, w_i^{(t)}) \in \left\{(\theta^{(t)}, y^{(t)}, w^{(t)}) \mid y^{(t)} \in \acceptancereg{\epsilon_{k+1}}{\epsilon_{k}} \right\}$ be a tuple that defines a particle from ABC posterior at iteration $t \leq k < T$, that is within a partitioned acceptance region band $\acceptancereg{\epsilon_{k+1}}{\epsilon_k}$. 
We seek a transition kernel
\begin{align}
    K_k(\theta^{(t+1)}\mid \theta^{(t)}_i) = \normal{\theta^{(t+1)}\mid \theta^{(t)}, \Sigma_i^{(k)}}
\end{align}
where covariance $\Sigma_i^{(k)}$ is optimal for mutating the particle $\theta^{(t)}_i$ associated with $\acceptancereg{\epsilon_{k+1}}{\epsilon_{k}}$ so that it would produce an observation from acceptance region $\acceptancereg{0}{\epsilon_{k+1}}$. The idea here is that the particles from each iteration $t$ that are associated with acceptance region $\acceptancereg{\epsilon_{k+1}}{\epsilon_{k}}$ would the transferred to a posterior that is defined by $\acceptancereg{0}{\epsilon_{k+1}}$ that is one step closer to the final acceptance region $\acceptancereg{0}{\epsilon_{T}}$ as defined by the fixed threshold sequence. Traditionally, all particles would be transferred to approximate a posterior defined by $\acceptancereg{0}{\epsilon_{t+1}}$, regardless of the associated values $y_i^{(t)}$.
The proposed modification can be achieved using the same approach as deriving locally optimal transition kernel covariance \eqref{eq:locally_optimal_cov}.
Now proposed ABC SMC joint proposal distribution for the tranformation at $t$th iteration  is of the form
\begin{align}\label{eq:strat_ABC_proposal}
    & \jointproposal{k}\left(\theta^{{(t)}}, \theta^{(t+1)}\right) \\\nonumber & =\frac{\ABC{\acceptancereg{\epsilon_{k+1}}{\epsilon_k}}{\theta^{(t)}}K_k(\theta^{(t+1)} | \theta^{(t)})  \prob{y^\prime \in \acceptancereg{0}{\epsilon_{k+1}} \mid \theta^{(t+1)}}}{\int \int \ABC{\acceptancereg{\epsilon_{k+1}}{\epsilon_k}}{\theta^{(t)}}K_k(\theta^{(t+1)} | \theta^{(t)}) \prob{y^\prime \in \acceptancereg{0}{\epsilon_{k+1}} \mid \theta^{(t+1)}} \intd \theta^{(t+1)}\intd \theta^{(t)}}.
\end{align}

\begin{figure}[ht!]
   \centering
   \includegraphics[trim={0 0 0 0}, clip, width = 1.0\textwidth]{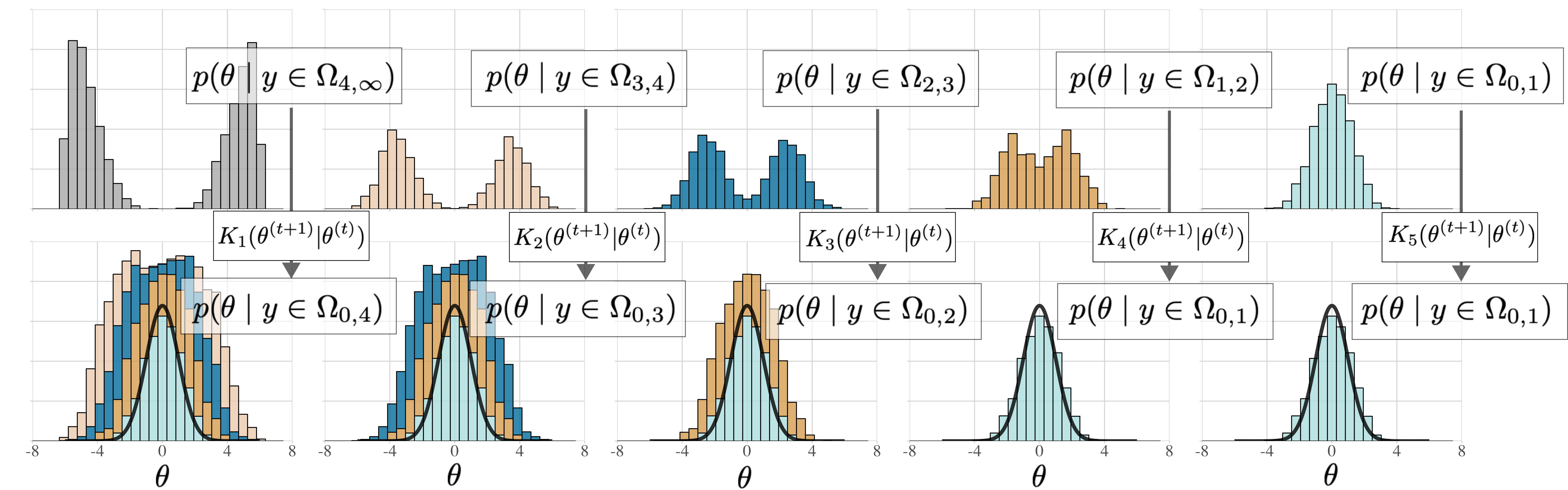}
   \caption{Example 1 continued. The particles residing in different partitioned components of the ABC posterior are being transitioned with a kernel defined by the cascading structure of improving ABC posterior estimates.}
   \label{Fig:transition_kernels}
\end{figure}

As before, we derive an optimal Gaussian kernel \eqref{eq:gaussian_kernel} by minimizing the Kullback-Leibler (KL) discrepancy between \eqref{eq:ABC_proposal} and
\begin{align}\label{eq:stratified_optimal_product}
\optimaljointproposal{k}\left(\theta^{(t)}, \theta^{(t+1)}\right) = \ABC{\acceptancereg{\epsilon_{k+1}}{\epsilon_{k}}}{\theta^{(t)}} 
p \left( \theta^{(t+1)} \mid y^\prime \in \acceptancereg{0}{\epsilon_{k+1}} \right),
\end{align}
which leads to
\begin{align}\label{eq:stratified_optimal_kernel_covariance}
    \Sigma^{(k)} = \int \! \! \int & (\theta^{(t)} \!-\! \theta^{(t+1)})(\theta^{(t)} \!-\! \theta^{(t+1)})^T \times \nonumber \\ & \times \ABC{\acceptancereg{\epsilon_{k+1}}{\epsilon_{k}}}{\theta^{(t)}} p \left( \theta^{(t+1)} \mid y^\prime \in \acceptancereg{0}{\epsilon_{k+1}} \right) \intd \theta^{(t)} \intd \theta^{(t+1)},
\end{align}
for which a locally optimal version at $\theta^{(t)}_{i}$ is
\begin{align}\label{eq:stratified_locally_optimal_kernel_covariance}
\Sigma_{i}^{(k)} &= \int \left( \theta_i^{(t)} - \theta^{(t+1)} \right)\left( \theta_i^{(t)}  - \theta^{(t+1)} \right)^T\ABC{\acceptancereg{0}{\epsilon_{k+1}}}{\theta^{(t+1)}} \intd \theta,
\end{align}
where $\theta_i^{(t)}$ is such that the associated observation $y_i^{(t)} \in \acceptancereg{\epsilon_{k+1}}{\epsilon_{k}}$.
In Figure \ref{Fig:transition_kernels} we illustrate the cascading nature of the ABC posterior approximations at a single iteration $t$ of the methods and how the transition kernels are locally optimally defined for each of the particles separately based on posterior component they belong to.

We express the average acceptance rate at the transition from round $t \rightarrow t+1$  \eqref{eq:average_acceptance_rate} as 
\begin{align}
    & \int \int \sum_{l=t+1}^T \sum_{k=t}^T \prob{y^\prime \in \acceptancereg{\epsilon_{l+1}}{\epsilon_{l}} \mid \theta^{(t+1)}} K_k(\theta^{(t+1)} | \theta^{(t)}) \ABC{\acceptancereg{\epsilon_{k+1}}{\epsilon_k}}{\theta^{(t)}} \intd \theta^{(t)} \intd \theta^{(t+1)} \nonumber \\
    &= \sum_{l=t+1}^T \sum_{k=t}^T \int \int \prob{y^\prime \in \acceptancereg{\epsilon_{l+1}}{\epsilon_{l}} \mid \theta^{(t+1)}} K_k(\theta^{(t+1)} | \theta^{(t)}) \ABC{\acceptancereg{\epsilon_{k+1}}{\epsilon_k}}{\theta^{(t)}} \intd \theta^{(t)} \intd \theta^{(t+1)} \nonumber  \\
    &= \sum_{l=t+1}^T \sum_{k=t}^T C_{l,k}^{(t)},
\end{align}
where $C_{l,k}^{(t)}$ is the average rate of transitioning a particle from posterior $p(\theta^{(t)} \mid y \in \acceptancereg{\epsilon_{k+1}}{\epsilon_k})$ with kernel $K_k(\cdot \mid \cdot)$ and generating a synthetic observation $y \in \acceptancereg{\epsilon_{l+1}}{\epsilon_l}$. 
Given that we are approximating $C_{l,k}^{(t)} \approx C_{l,k}, l = t+1, \ldots, T, k = t, \ldots, T $ at each iteration of the ABC SMC, we can use these regional weights as a basis for a stratification strategy to compose an importance sampling distribution for more efficient simulation acceptance, by using the information we have obtained about the different  different partitioned posterior components and their average acceptance rates $C_{l,k}$.
In practice we approximate $C_{l,k}$ by counting the frequencies $f_{l,k}^{(t)}$ how often samples drawn from 
\begin{align*}
p(\theta^{(t+1,t)} \mid y \in \acceptancereg{\epsilon_{k+1}}{\epsilon_{k}}):= \int K_k(\theta^{(t+1)} \mid \theta^{(t)})p(\theta^{(t)} \mid y \in \acceptancereg{\epsilon_{k+1}}{\epsilon_k}) \intd \theta^{(t)}
\end{align*}
generate observations $y \in \acceptancereg{\epsilon_{l+1}}{\epsilon_l}$. This is carried out as a part of the weighted ABC sampling at each iteration. The distribution of the posterior predictive observations from the posterior at $t$th iteration that we will denote here as  $y^{(t+1, t)}$ is 
\begin{align*}
p(y^{(t+1, t)} \mid y \in \acceptancereg{\epsilon_{k+1}}{\epsilon_{k}}) := \int p(y^{(t+1, t)} \mid \theta^{(t+1, t)} )p(\theta^{(t+1,t)} \mid y \in \acceptancereg{\epsilon_{k+1}}{\epsilon_{k}}) \intd \theta^{(t+1, t)}
\end{align*}
can be estimated with the probability mass distribution of the discretised events $y^\prime \in \acceptancereg{\epsilon_{l+1}}{\epsilon_{l}} | y \in \acceptancereg{\epsilon_{k+1}}{\epsilon_{k}}$. This is approximated using the frequencies
\begin{align}\label{eq:pred_prob_component}
    \widehat{C}_{l,k}^{(t)} = \sum_{m=1}^{t} \frac{f_{l,k}^{(m)}}{\sum_{m^\prime=1}^{t} \sum_{l^\prime = 1}^T  f_{l^\prime,k}^{(m^\prime)} }, \quad l = 1, \ldots, T, \quad k = t, \ldots, T.
\end{align}

\begin{figure}[ht!]
   \centering
   \includegraphics[trim={0 0 0 0}, clip, width = 0.8\textwidth]{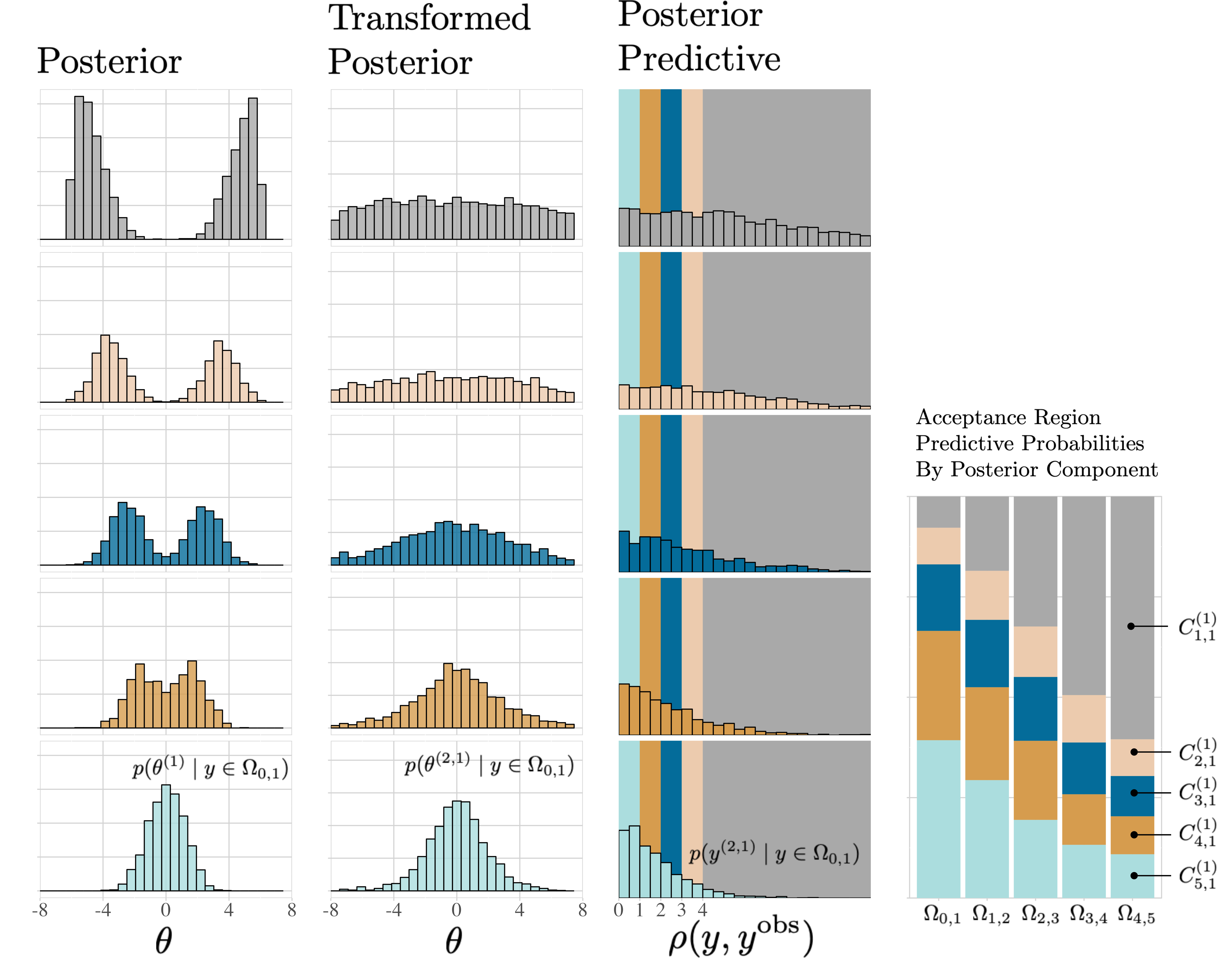}
   \caption{Example 1 continued. After the first ABC SMC iteration the particles are mutated and new set of synthetic observations are generated. We calculate the probabilities for each posterior component of generating observations in each of the acceptance regions. These probabilities are cumulatively improved along with the iterations. In this case we observe that as expected, posterior component related to $\Omega_{0,1}$ has the largest probabilities of generating observations that will be accepted in any round. These component-wise predictive probabilities are then used to re-weight the ABC posterior sample weights.}
   \label{Fig:pred_weighting}
\end{figure} 

The full weight for iteration $t$ is obtained by first taking the sum over different acceptance regions that are within $\acceptancereg{0}{\epsilon_{t+1}}$ 
\begin{align}\label{eq:strataweight}
    \widehat{W}_k^{(t)} = \sum_{l=t+1}^{T} \widehat{C}_{l, k}^{(t)}  = \sum_{l=t+1}^{T} \sum_{m=1}^{t} \frac{f_{l,k}^{(m)}}{\sum_{m^\prime=1}^{t} \sum_{l^\prime = 1}^T  f_{l^\prime,k}^{(m^\prime)} }, \quad k = t, \ldots, T.
\end{align}
We illustrate the cumulative predictive probability weighting in Figure \ref{Fig:pred_weighting}.
Cumulative predictive probabilities improve along with the iterations as we obtain realisations of posterior predictive observations from different posterior components. We wish to re-balance the importance sampling distribution by giving more weight to particles that have been observed to generate samples within acceptance regions that will be included in the subsequent iterations of the algorithm.
At iteration $t$, our approximation to the posterior importance sampling distribution can be constructed from the posterior distribution  $\{(\theta^{(t)}_i, y^{(t)}_i, w_i^{(t)}, I^{(t)}_i\}_{i=1}^N$, where $I^{(t)}_i \in \{1, \ldots, T\}$ indicates the corresponding partition of the observation space.  For example, if $I^{(t)}_i = k$, then $y^{(t)}_{i} \in \acceptancereg{\epsilon_{k+1}}{k}$. The empirical importance sampling distribution is $\{( \theta^{(t)}_{i}, y^{(t)}_{i}, \hat{w}^{(t)}_{i}, I^{(t)}_i )  \}_{i=1}^N$, where $\hat{w}^{(t)}_{i}$ is the re-balanced weight. It is calculated using the stratified observation space by first normalizing the weights within each strata, then weighting them with \eqref{eq:strataweight} and finally re-normalising the whole set, as
\begin{align}\label{eq:reweighted_abc_sample}
    \hat{w}^{(t)}_{i} \propto \frac{w_{i}^{(t)}}{\sum_{j: I^{(t)}_j =I^{(t)}_i}  w_{j}^{(t)}} \cdot \widehat{W}_{I_i^{(t)}}^{(t)}.
\end{align}
Based on the re-balanced weights we draw on average $\widehat{W}_{k}^{(t)}$ of the samples from $\{(\theta^{(t)}_i, y^{(t)}_i, w_i^{(t)}, I^{(t)}_i) : I_i^{(t)} = k\}_{i=1}^N$ because the predictive distribution conditioned on $y \in \acceptancereg{\epsilon_{k+1}}{\epsilon_{k}}$ have been approximated empirically to generate $\widehat{W}_{k}^{(t)}$ of the samples within $\acceptancereg{0}{\epsilon_{t+1}}$. This stratified sampling approach should accelerate the acceptance rate especially in the early iterations of ABC SMC when the transition kernels can have large covariances. The largest weights for the samples calculated via \eqref{eq:abcsmcweight} are assigned to the samples from the tails of the importance sampling distribution, when the corresponding synthetic observations are within the acceptance region. However, these outlying samples with large weights may also result in lower acceptance rate on the subsequent iterations.

We can estimate the locally optimal covariance kernel \eqref{eq:stratified_locally_optimal_kernel_covariance} based on the weighted sample by 
\begin{align*}
\hat{\Sigma}_{i}^{(k)} &= \sum_{j=1}^{N^\prime} \hat{\omega}^{(t)}_{I(j)}\left( \theta_i^{(t)} - \theta_{I(j)}^{(t)} \right)\left( \theta_i^{(t)}  - \theta_{I(j)}^{(t)} \right)^T,
\end{align*}
where $\theta^{(t)}_i$ is associated to acceptance region $\acceptancereg{\epsilon_{k+1}}{\epsilon_k}$. The index set $\{I(1), \ldots, I(N^\prime)\} \subset \{1, \ldots, N\}$ is such that $I^{(t)}_{I(i)} \geq k+1$, and the re-normalised weights are calculated as
\begin{align}
    \hat{\omega}^{(t+1)}_{I(i)} = \frac{\hat{w}_{I(i)}^{(t)}}{\sum_{i=1}^{N^\prime} \hat{w}_{I(i)}^{(t)}}.
\end{align}
The proposed algorithm is summarised for a single iteration of ABC SMC in Algorithm \ref{alg:postpredbinning}.

\begin{algorithm}[!ht]
\caption{Empirical estimation of posterior predictive probability of acceptance events}
\begin{algorithmic}[1]
\State \textbf{Input 1:} Posterior sample $\{ ( \theta^{(t)}_{i}, y^{(t)}_{i}, w^{(t)}_{i}, I_i^{(t)} ) \}_{i=1}^{N}$.
\State \textbf{Input 2:} Frequencies $f_{l,k}^{(m)}, m=1,\ldots, t, k= 1, \ldots, T, l = 1, \ldots, T$ as in \eqref{eq:pred_prob_component}.
\State Calculate importance sampling distribution $\{ ( \theta^{(t)}_{i}, y^{(t)}_{i}, \hat{w}^{(t)}_{i}, I_i^{(t)} )  \}_{i=1}^{N}$ using \eqref{eq:reweighted_abc_sample}
\State Set $n=1$
\State Set $f_{l, k}^{(t+1)} = 0, k= 1, \ldots, T, l = 1, \ldots, T$
\While{$n  \leq N$}
  \State Select $(\theta^{(t)}, y^{(t)}, I^{(t)})$ from $\{ (\theta^{(t)}_{i}, y^{(t)}_i, \widehat{w}_i^{(t)}, I_i^{(t)}) \}_{i=1}^{N}$ with probability $\widehat{w}_i^{(t)}$
  \State Generate $\theta^\prime \thicksim K_{I^{(t)}}(\theta \mid \theta^{(t)} )$ 
  \State Generate $y^{(t+1, t)} \sim f(y \mid \theta^\prime)$
  \State Find $k$ such that $ y^{(t+1, t)} \in \acceptancereg{\epsilon_{k+1}}{\epsilon_{k}}$
  \State Set $f_{k, I^{(t)}}^{(t+1)} = f_{k, I^{(t)}}^{(t+1)} + 1$
  \If {$y^{(t+1, t)} \in \acceptancereg{0}{\epsilon_{t+1}}$}
    \State $\theta_n^{(t+1)} = \theta^\prime$
    \State $y_n^{(t+1)} = y^{(t+1, t)}$
    \State $I^{(t+1)}_n = k$
    \State Calculate $w_n^{(t)} \propto p(\theta_n) / \sum_{i=1}^N \widehat{w}_i^{(t)} K_{I^{(t)}}(\theta^\prime \mid \theta^{(t)}_i)$
    \State $n = n + 1$
  \EndIf
\EndWhile
\State Normalise $w_n^{(t)} = w^{(t)}_n / \sum_{i=1}^N w^{(t)}_i$
\State \textbf{Output 1:} Posterior sample $\{ ( \theta^{(t+1)}_{i}, w^{(t+1)}_{i}, y^{(t+1)}_{i}, I_i^{(t+1)} ) \}_{i=1}^{N}$
\State \textbf{Output 2:} Frequencies $f_{l,k}^{(m)}, m=1,\ldots,t+1, k= 1, \ldots, T, l = 1, \ldots, T$
\end{algorithmic} \label{alg:postpredbinning}
\end{algorithm}

\section{Stopping rule}\label{sec:stopping_rule}

ABC SMC is an iterative sampling algorithm that will require more computational resources after each iteration as the acceptance threshold is decreased, either adaptively or following a predetermined schedule, as in the case of the approach presented here. 
However, after some point the improvements we are obtaining from decreasing the threshold may not be significant enough to justify the required resources for improving the approximation. 
We propose a novel stopping rule based on the evolution of the prediction probabilities $C_{l,k}^{(t)}$.
The prediction probabilities of the partitioned acceptance regions can be expressed as a probability mass function defined by the probabilities
\begin{align}\label{eq:predictive_pmf_acc_region}
    \prob{y^{(t)} \in \acceptancereg{\epsilon_{l+1}}{\epsilon_{l}} \mid y \in \acceptancereg{\epsilon_{k+1}}{\epsilon_{k}}} &= C_{l,k}^{(t)}, \quad l=1,\ldots, T.
\end{align}
The target of the ABC SMC is the approximate posterior at the final iteration $T$ and we use the probability mass $C_{:,T}^{(t)}$ as a proxy to monitor the evolution of the approximate posterior. We use notation $C_{:,k}^{(t)}$ to denote the probability mass function of the acceptance region events. If at the iteration $t$ the approximate $\widehat{C}_{:,t}^{(t)}$ is already resembling $\widehat{C}_{:,T}^{(t)}$, we can consider stopping the algorithm early. Computationally this convenient as we can count the frequencies \eqref{eq:pred_prob_component} as part of the proposed  Algorithm \ref{alg:postpredbinning}.
We can use the Kullback-Leibler (KL) divergence
\begin{align}\label{eq:KL_div}
    \KL{C_{:,T}^{(t)}}{C_{:,t}^{(t)}} &= \sum_{l=1}^T C_{l,T}^{(t)} \log \left( \frac{C_{l,T}^{(t)}}{C_{l,t}^{(t)}} \right)
\end{align} 
as the quantified difference between the target and the current state. Note that by the definition we can start monitoring the convergence only after the second iteration as all the parameters on the first round are from the prior instead of the components of the partitioned approximate posterior.
It is also notable that the early stopping rule monitors whether we have achieved posterior predictive performance of the final iteration round. Therefore if the final threshold is very small, it is possible that our empirical approximation of $C_{:, T}^{(t)}$ is poor, or in the worst case, does not contain a single sample. It is recommended to set up a minimum number of samples that are required for approximating $C_{:, T}^{(t)}$, and only after it is achieved, the early stopping rule monitoring is activated.

\begin{figure}[ht!]
   \centering
   \includegraphics[trim={0 0 0 0}, clip, width = 0.6\textwidth]{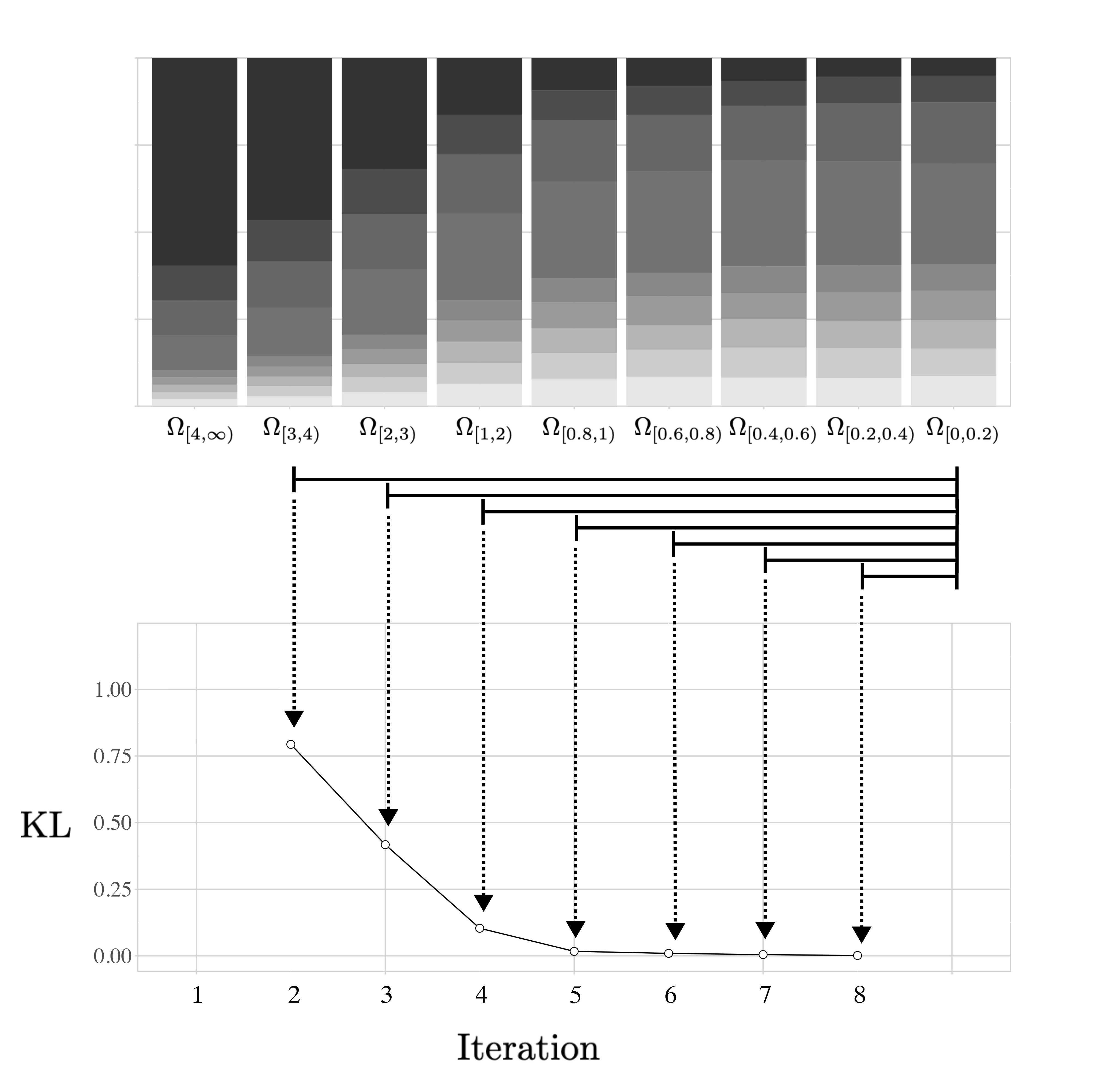}
   \caption{Example 1 continued. Upper figure bars illustrate the predicted probability mass functions (pmf) \eqref{eq:predictive_pmf_acc_region}. At each iteration we calculate pmf defined by \eqref{eq:predictive_pmf_acc_region} and the corresponding KL divergence \eqref{eq:KL_div} that is reported on the bottom figure. }
   \label{Fig:KL_divergence}
\end{figure}

\begin{exmp}
    We illustrate the behaviour of the KL divergence using the model from Example \ref{ex:1}  and augmenting the model with more acceptance regions that are defined by threshold sequence $\epsilon_{1:9} = [\infty, 4,3,2,1,0.8,0.6,0.4,0.2]$. The experiment is carried out with $N=150000$ samples. We calculate the posterior predictive probabilities of the acceptance regions using the cumulative frequencies and for each of the iterations we calculate the KL divergence between the consecutive probability mass functions. The evolution of KL is illustrated in Figure \ref{Fig:KL_divergence}. We see how the KL divergence plateaus after the $5$th iteration as the predicted probability mass function does not significantly change even when the threshold of the ABC posterior decreases.  
\end{exmp}

\section{Results}\label{sec:examples}

We compare the proposed stratified sampling ABC SMC approach to a globally and locally optimal ABC SMC methods introduced in Section \ref{sec:overview} denoted by \textit{Global} and \textit{Local}, respectively. Two version of the stratified sampling approach are tested here. \textit{Stratified simple} uses only the new transition kernel, whereas \textit{Stratified} used the transition kernel and the proposed importance sampling distribution. In each test case, we have fixed the unknown parameters $\theta_0$ and repeated the experiment $50$ times with different randomly generated observations $\yobs \thicksim p(y \mid \theta_0)$. 

We assume that accuracy-wise all the methods will perform similarly as the main difference of the methods is the sampling strategy. We investigate the acceptance rates, and estimation accuracies in three examples which have been widely used in ABC literature for benchmarking methods. In addition, we investigate the performance of the proposed method when solving a real problem of inferring transmission dynamics of \textit{Streptococcus pneumoniae} in day care centers. As estimation accuracy we report the medians of iteration-wise sample means in addition to the interval: median of the sample mean $\pm 1.96 $ $\cdot$ median of the sample standard deviation. In addition to accuracy we report the KL-divergence of the predicted acceptance region probabilities to investigate its usability as an early termination indicator. Acceptance rates, total number of generated samples and KL convergence monitoring quantity are reported as medians and interquartile ranges over the repetitions.

\begin{figure}[ht!]
   \centering
   \includegraphics[trim={0 0 0 0}, clip, width = \textwidth]{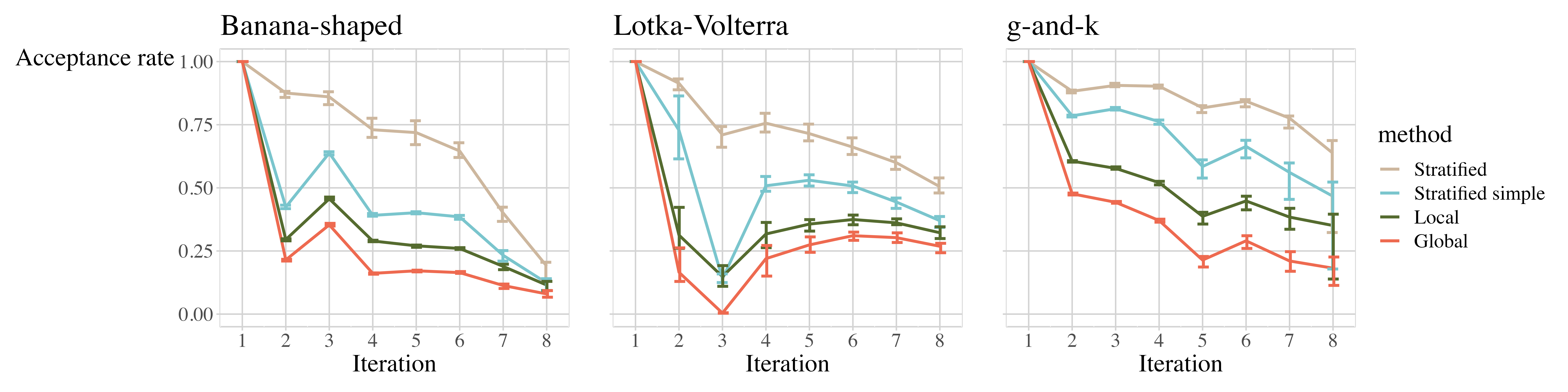}
   \caption{Acceptance rates by iteration round for the four compared methods in the three experiments with 2, 3 and 4 parameters as described in Sections \ref{ss:banana}
, \ref{ss:lotka_volterra} and \ref{ss:g_and_k}.  Acceptance rates are consistently higher for the proposed method in all experiments. The benefit of improved acceptance rate cumulates over the iteration rounds.} 
   \label{Fig:acceptance_rates}
\end{figure} 

\begin{figure}[ht!]
   \centering
   \includegraphics[trim={0 0 0 0}, clip, width = \textwidth]{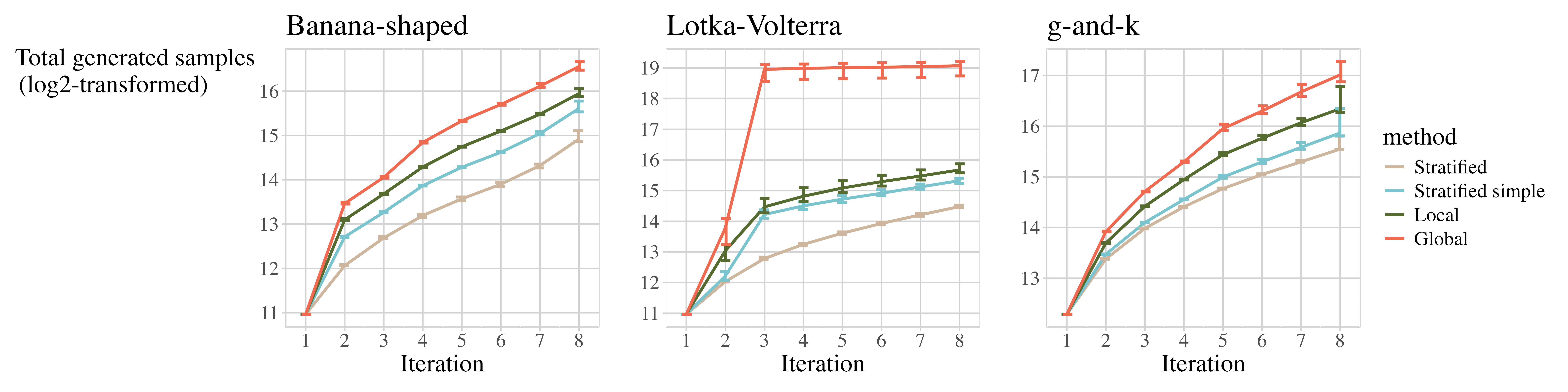}
   \caption{ Total cumulative count of generated samples by iteration round for the four compared methods in the three experiments with 2, 3 and 4 parameters as described in Sections \ref{ss:banana}
, \ref{ss:lotka_volterra} and \ref{ss:g_and_k}. The total cumulative number of samples indicate that the new stratified sampling method is more effective than the locally optimal ABC SMC. In banana shaped and G-and-k distribution examples, the locally optimal ABC SMC has reached sixth iteration, whereas the new proposed method is already on the final iteration round. On Lotka-Volterra model, when newly proposed method has reached the final iteration round, other methods have spent at least the same number of samples to reach the third iteration.  
}
   \label{Fig:generated_samples}
\end{figure} 

\subsection{Banana shaped distribution}\label{ss:banana}

\begin{figure}[htbp]
  \centering
   \includegraphics[trim={0 0 0 0}, clip, width = \textwidth]{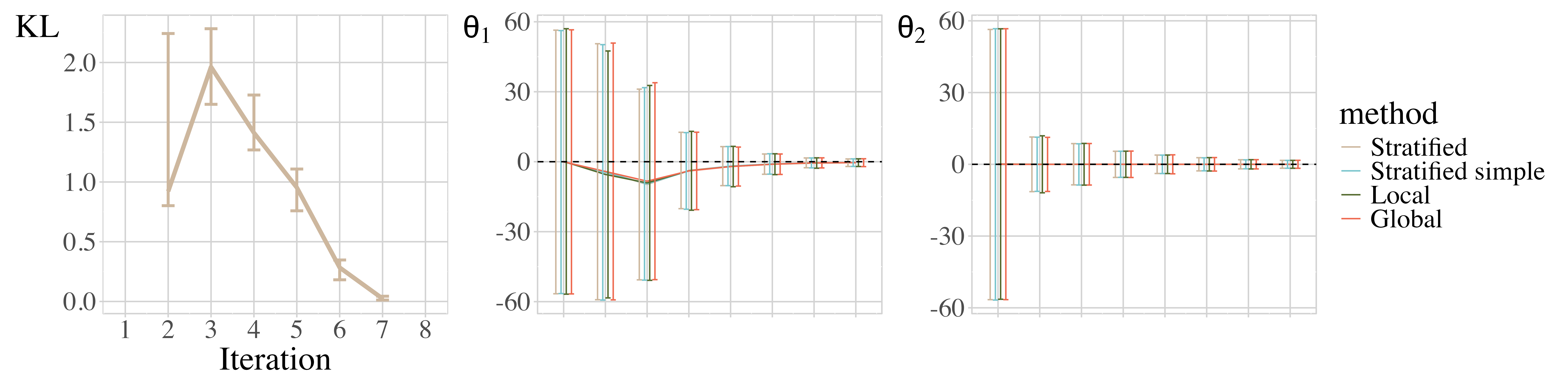}
   \caption{Evolution of the sample mean and $1.96\cdot$standard deviation and the early termination convergence diagnostic over $8$ iterations in Banana-shaped distribution example. True parameter values used to simulate all the datasets are marked with dashed lines ($\theta_1 = \theta_2 = 0$).  }
   \label{Fig:BS}
 \end{figure}

The first example is a two-parameter banana shaped posterior distribution, where the parameter dependence   structure is a challenge for the sampling methods. The model was introduced by \citet{haario1999} and used by e.g.~\citet{filippi2013} as a benchmarking example for ABC SMC. The likelihood and the prior are defined as 
\begin{align}
    p(y \mid \theta_1, \theta_2) &= \normal{\begin{bmatrix} \theta_1 \\ \theta_1 + \theta_2^2 \end{bmatrix}, \begin{bmatrix} 1 & 0 \\ 0 & 0.5 \end{bmatrix}} \\
    p(\theta_1, \theta_2) &= \unif{[-50, 50]}{\theta_1} \cdot \unif{[-50, 50]}{\theta_2}.
\end{align}
As divergence function we use Euclidean distance.
Methods use the threshold sequence $[\infty,$ $100,$ $50,$ $20,$ $10,$ $5,$ $2,$ $1]$, and at each iteration the posteriors are approximated with 2000 samples.  Acceptance rates provided by the proposed methods are significantly improved compared to local and global sampling strategies as seen in Figure \ref{Fig:acceptance_rates}. Even using only the proposed novel transition kernel (Stratified simple) the acceptance rate is better, but the stratified sampling strategy does improve the results markedly over almost all iterations. 
The estimation accuracy is slightly improved with the proposed methods and early termination convergence monitoring does indicate that the final iteration could be unnecessary. It is notable that the KL might be unreliable in the early iterations of the algorithm as indicated by the initial increase in the monitored divergence. 
The cumulative number of generated samples to achieve the required amount of samples is reported in Figure \ref{Fig:generated_samples}. Assuming the we would terminate the execution at the seventh round, we observe that e.g.~the locally optimal ABC SMC has reached only the fourth iteration round given the same simulation budget. If we compare the estimation errors reported in Figure \ref{Fig:BS} on fourth and seventh iteration rounds of locally optimal ABC SMC and stratified sampling ABC SMC, respectively, we notice that the proposed method is much more efficient.

\subsection{Lotka-Volterra}\label{ss:lotka_volterra}

Lotka-Volterra model is a stochastic kinetic system describing predator-prey dynamics. We adapt the implementation from  \citet{owen2015}.
The system is defined by three reactions R1, R2 and R3
\begin{align}
    & \text{Prey birth (R1): } X_1 \rightarrow 2 X_1 \\
    & \text{Prey death \& Predator birth (R2): } X_1 + X_2  \rightarrow 2 X_2 \\
    & \text{Predator death (R3): } X_2 \rightarrow 0.
\end{align}
We simplify our model by fixing the initial numbers of prey and predator populations to $100$ and $50$, respectively, and not adding observation noise to the state. 
The prior distributions for the hazard rates $r_i, \ i=1,2,3$ of the reactions are set as
\begin{align}
    \log(r_i) \thicksim \mathsf{Unif}(-6, 1), \quad i = 1,2,3
\end{align}
We use summarised data as observations and Euclidean distance function. The summaries are defined as
\begin{align*}
    S1 &= \text{Mean of prey population over the simulation} \\
    S2 &= \text{Mean of predator population over the simulation} \\
    S3 &= \text{Logarithm of variance of prey population over the simulation} \\
    S4 &= \text{Logarithm of variance of predator population over the simulation} \\
    S5 &= \text{Auto-correlation of lag 1 of prey population over the simulation} \\
    S6 &= \text{Auto-correlation of lag 1 of predator population over the simulation} \\
    S7 &= \text{Auto-correlation of lag 2 of prey population over the simulation} \\
    S8 &= \text{Auto-correlation of lag 2 of predator population over the simulation}
\end{align*}
All methods use a threshold sequence $[\infty,$ $200,$ $100,$ $90,$ $80,$ $70,$ $60,$ $50]$, and at each iteration the posteriors are approximated with 2000 samples.

\begin{figure}[htbp]
  \centering
   \includegraphics[trim={0 0 0 0}, clip, width = \textwidth]{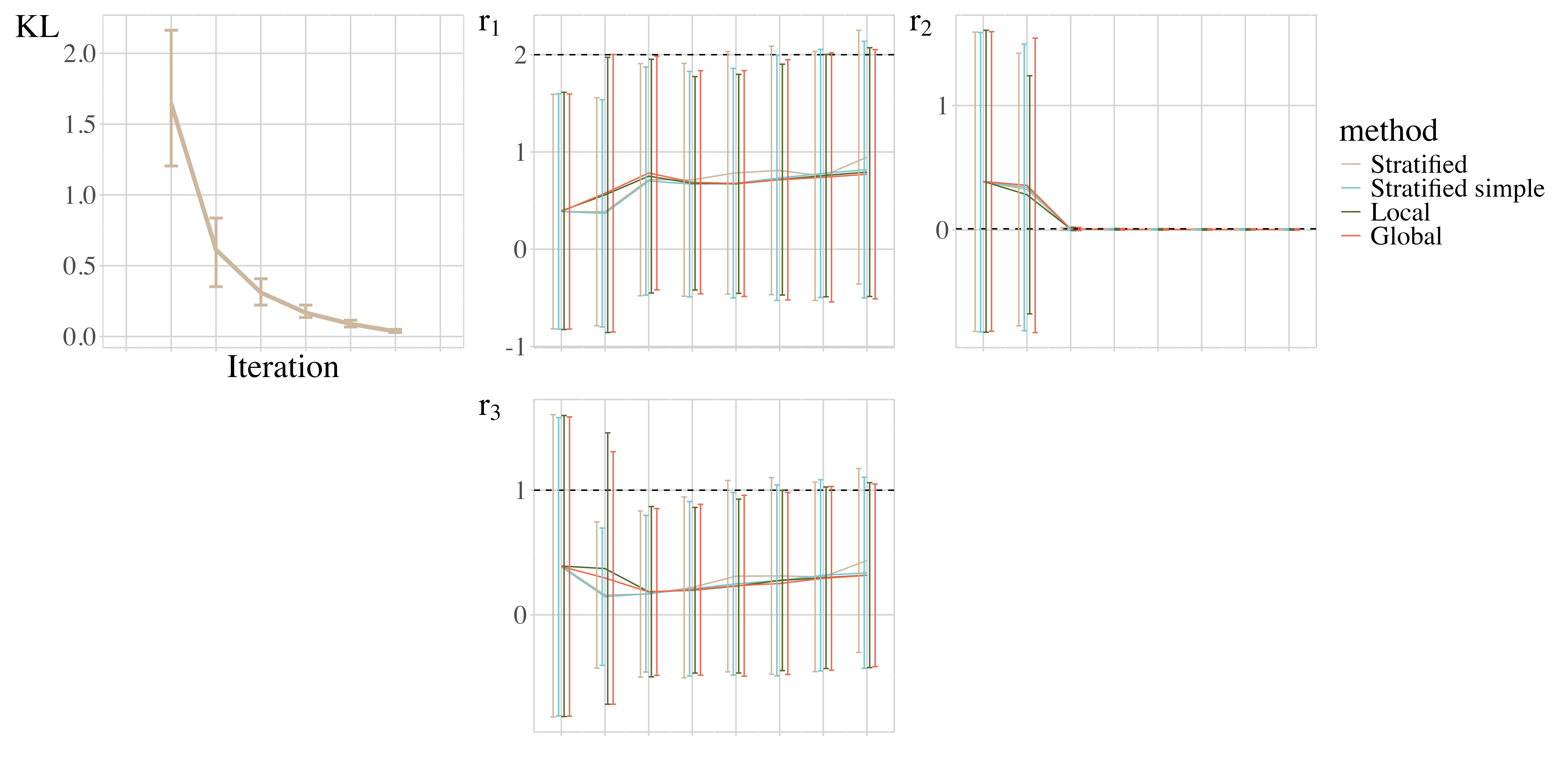}
   \caption{Evolution of the sample mean and $1.96\cdot$standard deviation and the early termination convergence diagnostic over $8$ iterations in Lotka-Volterra example. True parameter values used to simulate all observed datasets are marked with dashed horizontal lines ($r_1 = 2, r_2 = 0.01, r_3 = 1$).}
   \label{Fig:LV}
 \end{figure}

Acceptance rates are again significantly better with the proposed methods as reported in Figure \ref{Fig:acceptance_rates}. The estimation accuracy is better with the proposed methods, but the algorithms have not yet reached a stable state, but all methods should have improved performance when the threshold sequence would be continued as can be seen in Figure \ref{Fig:LV}. The total number of generated samples is less than a half with the stratified sampling ABC SMC than with the other methods on the final iteration round as reported on Figure \ref{Fig:generated_samples}.

\subsection{g-and-k-distribution}\label{ss:g_and_k}

\begin{figure}[htbp]
  \centering
   \includegraphics[trim={0 0 0 0}, clip, width = \textwidth]{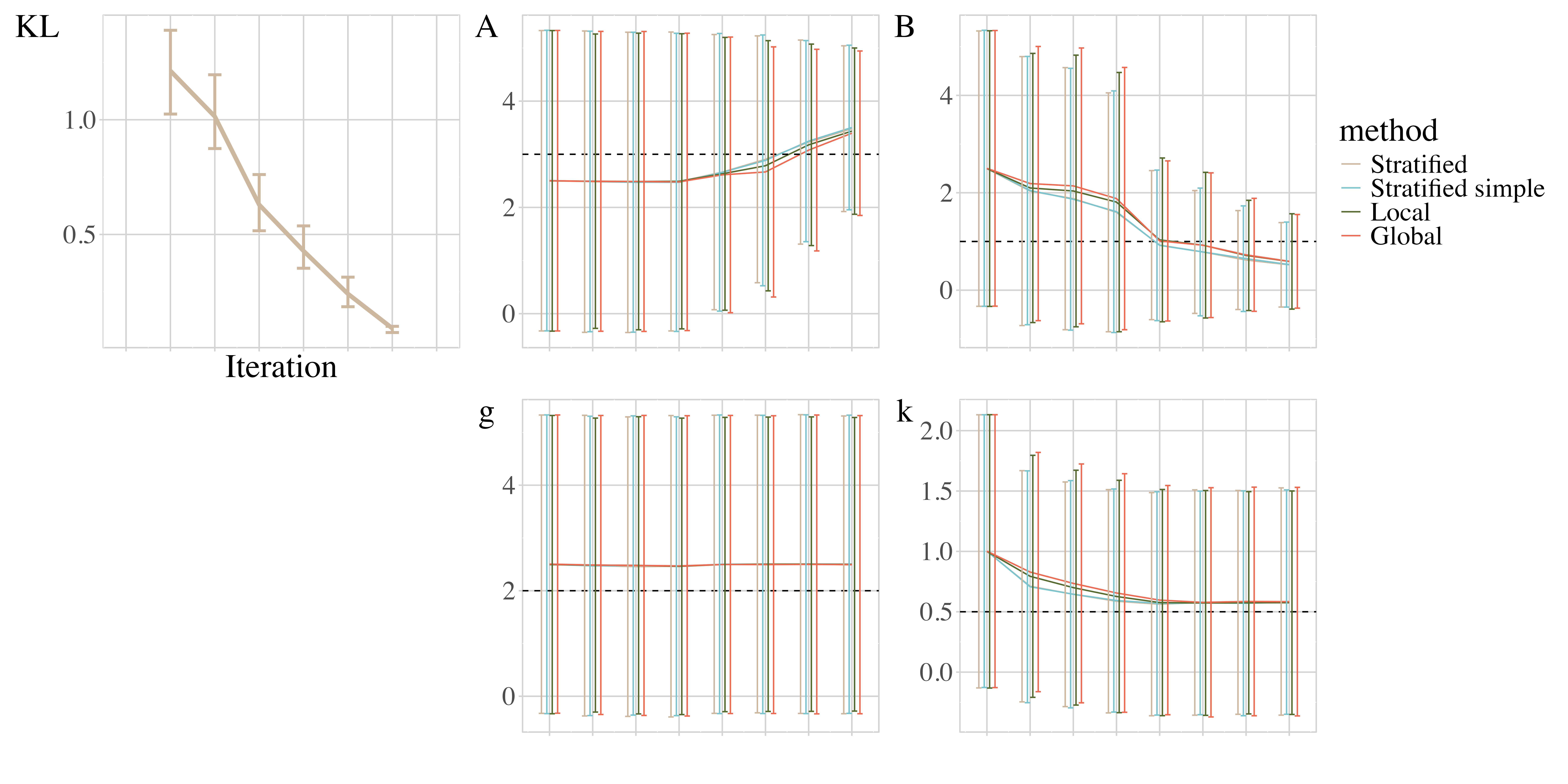}
   \caption{Evolution of the sample mean and $1.96\cdot$standard deviation and the early termination convergence diagnostic over $8$ iterations in g-and-k distribution example. True parameter values used to simulate all observed datasets are marked with dashed horizontal lines ($A = 3, B = 1, g = 2, k=0.5$).}
   \label{Fig:GNK}
 \end{figure}

The third test case we investigate is the univariate g-and-k distribution following \citep{drovandi2011}. 
We can draw a sample from the  distribution by first sampling $z(p)_i \stackrel{i.i.d}{\thicksim} \mathsf{Normal}(0,1), \quad i = 1, \ldots, 50$ and then drawing $y_i$ for each $i = 1, \ldots, 50$
\begin{align}
    y_i & \thicksim Q^\text{gk}(z(p)_i; A, B, g, k) \nonumber \\ &= A + B \cdot \left( 1 + 0.8 \cdot \frac{1 - \exp(-g \cdot z(p)_i)}{1 + \exp(-g \cdot z(p)_i)} \right) (1 + z(p)_i^2)^k z(p)_i.
\end{align}
We use uniform prior distributions for the parameters
\begin{align*}
    A, B, g &  \stackrel{\text{i.i.d}}{\thicksim} \mathsf{Unif}(0,5) \\
    k & \thicksim \mathsf{Unif}(0,2),
\end{align*}
and as the summary statistic we use the ordered sequence of observations
\begin{align*}
    [y_{I(1)}, y_{I(2)}, \ldots, y_{I(50)}], \quad i < j \Rightarrow y_{I(i)} \leq  y_{I(j)}.
\end{align*}
Euclidean distance is used as the divergence metric. All methods use a threshold sequence $[\infty,$ $100,$ $70,$ $50,$ $30,$ $27,$ $23,$ $20]$, and at each iteration the posteriors are approximated with 5000 samples. We observe similar results than on the previous experiments. Acceptance rates are significantly improved given the new proposed methods as illustrated in the Figure \ref{Fig:acceptance_rates}. From Figure \ref{Fig:GNK} we see that the estimation accuracy of the proposed method is a little better compared to local and global. The early termination quantity does not indicate convergence within the iterations which can also be seen the posterior mean estimate evolution. In Figure \ref{Fig:generated_samples} we see that the stratified method is a couple of iterations ahead simulation-budget-wise to locally optimal ABC SMC.

\section{Bacterial transmissions in day care centers}\label{ss:dcc}

We investigate a problem of estimating the transmission dynamics of \textit{Streptococcus pneumoniae} from strain prevalence data collected in day care units in Oslo metropolitan area. Originally, the transmission model parameters were solved using ABC SMC in the study by \citet{numminen2013}. Our formulation of the problem follows a slightly modified model and the set of summary statistics following \citep{gutmann2016bayesian}. We use a model where the number of the observed children equals the size of the center, i.e.~we don't simulate unobserved children. The data described in detail in \citet{vestrheim2008} consists of samples collected from 611 healthy children from 29 different day care centers. The 33 most prevalent strains were included in the data set.

The transmission dynamics are modeled separately for each day care center as a continuous time Markov process
\begin{align}
    & \prob{I_{is}(t + \delta t) = 1 \mid I_{is}(t) = 0} = \beta E_s(I(t)) + \Lambda P^s + o(\delta t), \text{ if } \sum_{j=1}^{N^s} I_{ij}(t) = 0, \\
    & \prob{I_{is}(t + \delta t) = 1 \mid I_{is}(t) = 0} = \theta(\beta E_s(I(t)) + \Lambda P^s + o(\delta t)), \text{ if } \sum_{j=1}^{N^s} I_{ij}(t) > 0, I_{is} = 0 \\
    & \prob{I_{is}(t + \delta t) = 0 \mid I_{is}(t) = 1} = \gamma + o(\delta t),
\end{align}
where $\beta \thicksim \mathsf{Unif}(0, 11), \Lambda \thicksim \mathsf{Unif}(0,2) $ and $ \theta \thicksim \mathsf{Unif}(0,1)$ are the internal infection parameter, the external infection parameter and the co-infection parameter, respectively. 
The first equation defines the probability for a child $i$ to be colonised with strain $s$ when the individual is not colonised with any of the other strains in the model, while the second equation described the probability when the child is already being colonised with at least one other strain. The last equation describes the recovery rate from colonization. We follow the formulation in \citet{numminen2013} and set $\gamma = 1$, and all other variables are scaled accordingly. The data is a single time point snapshot of the colonizations. We set the simulation time to $T=10$ time units, and use the state in the end as our observation. The resulting data consists of $N^\text{dcc}$ binary matrices $I_{is}^{(n)}, i = 1, \ldots, N^{c, \text{dcc}}, s = 1, \ldots, N^s, n=1, \ldots, N^\text{dcc}$, where $N^{c, \text{dcc}}$ is the number of children in the day care center $n$, $N^s$ is the number of bacterial strains and $N^\text{dcc}$ is the number of day care centers.

We use a set of four summary statistics for each of the day care centers, i.e.~Shannon index of of the strains, the number of different strains, the prevalence of carriage among children and the prevalence of more than one colonization among the children.  The summarised data of size 4 from 29 day care centers was first scaled based on the observed data so that the maximum of each of the summaries would be one. After the summary statistic scaling, the empirical cdfs of the joint set of 116 summaries was then compared via $L_1$-norm scaled with the coefficient 1/116.

\begin{figure}[htbp]
  \centering
   \includegraphics[trim={0 0 0 0}, clip, width = \textwidth]{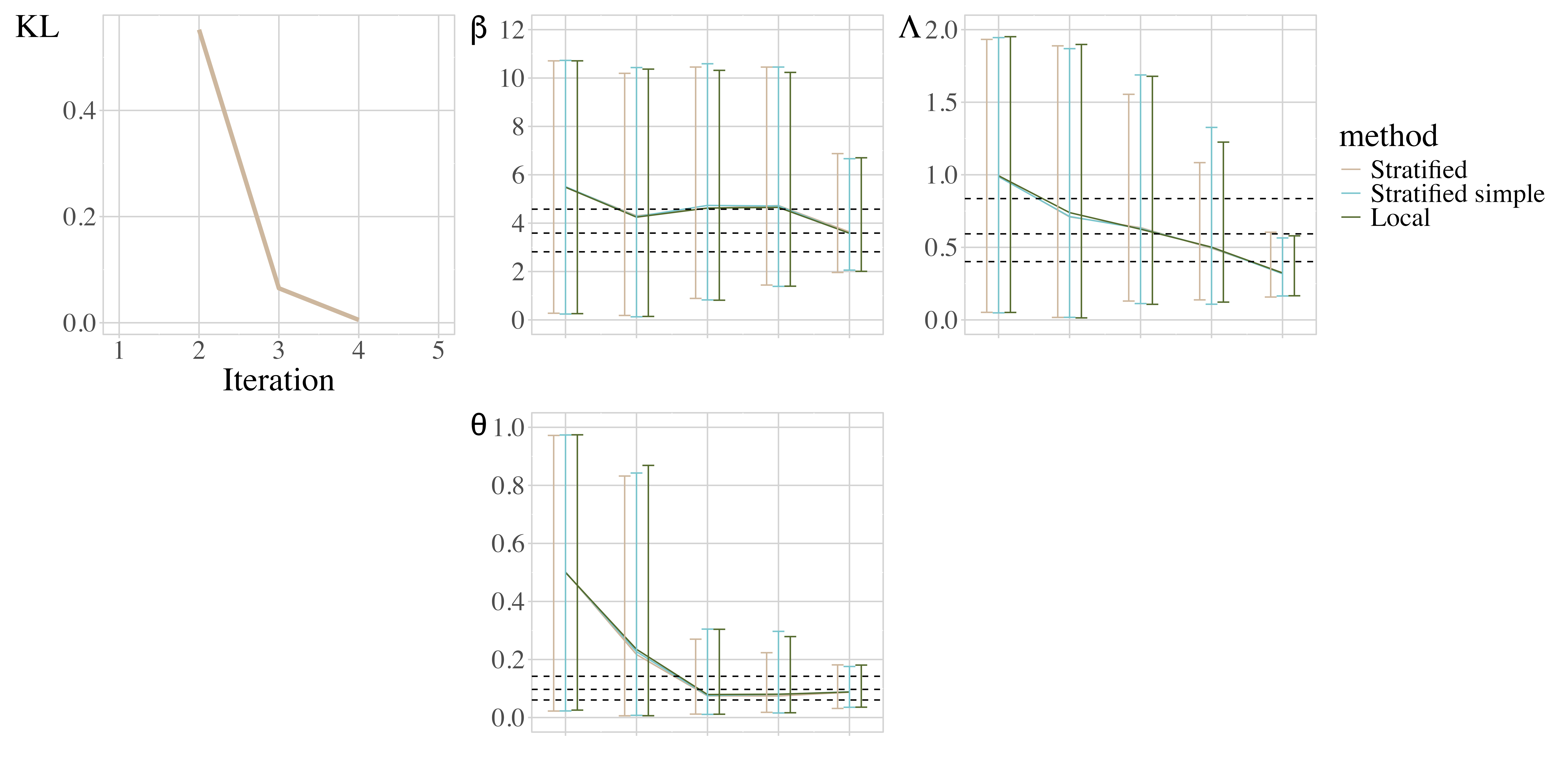}
   \caption{Evolution of the sample mean and the $95\%$-credible interval of the three parameters of the day care center model described in Section \ref{ss:dcc}. The horizontal dashed lines are the sample mean and $95\%$-credible intervals as reported in \citet{numminen2013}.}
   \label{Fig:dcc2}
 \end{figure}

\begin{figure}[htbp]
  \centering
   \includegraphics[trim={0 0 0 0}, clip, width = 0.8\textwidth]{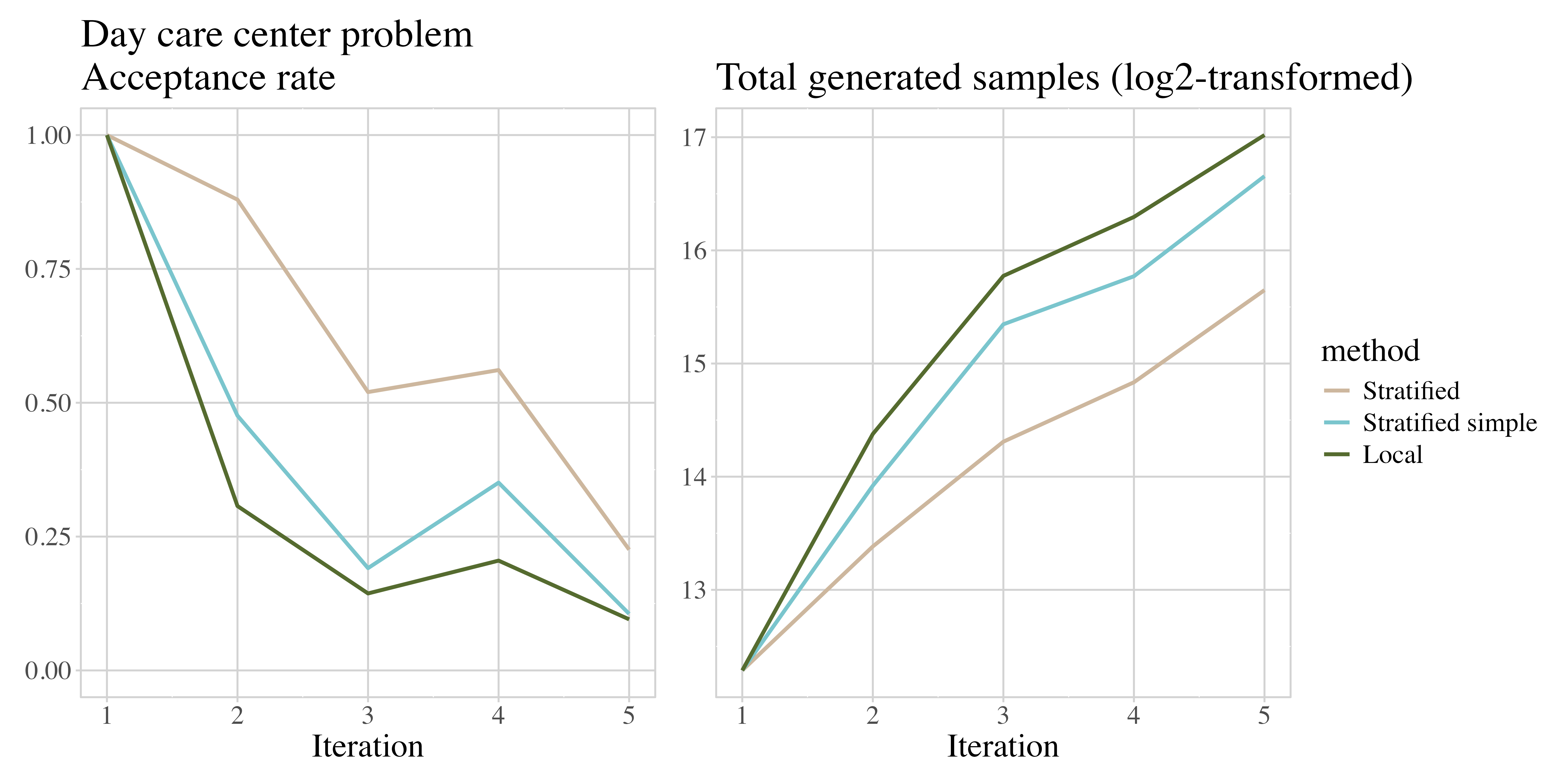}
   \caption{ Acceptance rates and the total number of generated samples by iteration round for the three compared methods in the day care center model described in Section \ref{ss:dcc}. Acceptance rates are consistently higher for the proposed method in all experiments. The benefit of improved acceptance rate cumulates over the iteration rounds, and eventually the new proposed method requires only half of the simulator queries to reach the final approximate posterior.}
   \label{Fig:dcc1}
 \end{figure}

The two versions of the stratified sampling approach are tested along with the locally optimal ABC SMC to infer the model parameters with $5000$ approximate posterior samples using a threshold sequence $[\infty, 0.5, 0.1, 0.08, 0.05]$. The results illustrated in Figure \ref{Fig:dcc2} indicate that there are no significant differences in the inferred approximate posterior. Using the proposed algorithm, we do obtain the final posteriors by using less than half of the simulations when comparing the stratified and the locally optimal versions in Figure \ref{Fig:dcc1}. 
There is a discrepancy between the posterior inferred by \citet{numminen2013} and our approximate posteriors in the estimate of the parameter $\Lambda$ that is most likely caused by the difference in the used models, as described earlier in the section.

\section{Conclusions}\label{sec:conclusion}

Here we introduced an approach to improve the acceptance rate of ABC SMC using a stratification of the acceptance region. Using the accumulated information of the posterior predictive distributions, we can re-weight the importance sampling distribution of the ABC SMC to draw particles that are more likely to produce observations that will be accepted at the subsequent iterations of the algorithm. In addition we described a transition kernel that takes the cascading formulation of the algorithm into account, and a novel early termination diagnostic based on the KL divergence of the predictive distribution of the acceptance regions. 

Our simulations confirm feasibility of the proposed method and demonstrate that it achieves significantly improved acceptance rate than the currently most popular versions of the ABC SMC algorithm. Even if there are no large differences between the inference methods in estimation accuracy when considering specific iteration rounds, the improved acceptance rate means that the newly proposed method achieves the accuracy levels earlier. A restriction of the method, however, is the requirement of determining the threshold sequence beforehand. In practice this means that it is necessary to obtain problem domain specific knowledge about the behavior of the discrepancy measure using explorative analyses of the simulator output with feasible inputs. Of note, many such pre-inference analyses would also match the good practices from the traditional Bayesian workflow \citep{gelman2020}. In situations where the \textit{a priori} analyses would not be considered feasible, it would be necessary to resort to adaptive algorithms not necessitating pre-specified thresholds. Many adaptive ABC SMC algorithms have been designed to effectively skip unnecessary evaluations of ABC posteriors by finding the threshold values based on the current posterior sample. Even if these methods are not designed to improve the acceptance rate, they can achieve good performance with less total evaluations by skipping unnecessary steps when using fixed threshold sequencing. Further work could thus be done to combine more efficient sampling with adaptation of the acceptance regions to the data.

\section*{Acknowledgements}

This research was supported by Research Council of Norway, grant no. 299941 and through its Centre of Excellence Integreat - The Norwegian Centre for knowledge-driven machine learning, project number 332645.

\bibliographystyle{abbrvnat}
\bibliography{references}

\end{document}